\documentclass[12pt]{iopart}

\usepackage{graphicx}
\usepackage[usenames,dvipsnames]{color}

\expandafter\let\csname equation*\endcsname\relax
\expandafter\let\csname endequation*\endcsname\relax
\usepackage{amsmath}
\usepackage{amssymb}

\usepackage{textcomp}

\newcommand{\karman}{von~K\'arm\'an}

\newcommand{\eg}{\emph{e.g.}} 
\newcommand{\ie}{\emph{i.e.}} 
\newcommand{\fig}[1]{Fig.~\ref{#1}} 

\let\di=\displaystyle

\begin{document}

\title{{Large spheres motion in a non homogeneous turbulent flow}}

\author{{Nathana\"el Machicoane$^{1,3}$, Robert Zimmermann$^{1,3}$, Lionel Fiabane$^{1,3}$, Micka\"el Bourgoin$^2$, Jean-Fran\c{c}ois Pinton$^1$, and Romain Volk$^{1}$}\footnote{Author to whom correspondence should be addressed.}}

\address{$^1$~Laboratoire de Physique de l'\'Ecole Normale Sup\'erieure de Lyon, CNRS UMR 5672, 46 All\'ee d'Italie, F-69007 Lyon, France}
\address{$^2$~Laboratoire des \'Ecoulements G\'eophysiques et Industriels, CNRS/UJF/G-INP UMR 5519, BP53, F-38041 Grenoble, France}
\address{$^3$~These authors contributed equally to this work.}
\ead{romain.volk@ens-lyon.fr}

\begin{abstract}
We investigate the dynamics of very large particles freely advected in a turbulent \karman{} flow. Contrary to other experiments for which the particle dynamics is generally studied near the geometrical center of the flow, we track the particles in the whole experiment volume. We observe a strong influence of the mean structure of the flow that generates an unexpected large-scale sampling effect for the larger particles studied; contrary to neutrally buoyant particles of smaller yet finite sizes that exhibit no preferential concentration in homogeneous and isotropic turbulence (Fiabane~\emph{et al.}, Phys.~Rev.~E {\bf 86}(3), 2012). We find that particles whose diameter approaches the flow integral length scale explore the \karman{} flow non-uniformly, with a higher probability to move in the vicinity of two tori situated near the poloidal neutral lines. This preferential sampling is quite robust with respect to changes of any varied parameters: Reynolds number, particle density and particle surface roughness. 
\end{abstract}
\submitto{\NJP}
\maketitle


\section{Introduction}\label{sec:intro}
Solving the dynamics of a sphere of diameter $d$ freely advected in a turbulent flow requires a knowledge of all forces acting on it.  The determination of these forces is a long-standing challenge. It has been solved to a large extent for particles much smaller than the dissipation (or Kolmogorov) length scale $\eta$~\cite{Gatignol:1983, Maxey:pof1983}. In this case, the equation governing the particle velocity $\mathbf{v}$ is determined once the fluid velocity $\mathbf{u}$ is known, as forces are dominated by Stokes drag and added mass effects~\cite{Elghobashi:jfm1992}. Very small neutrally buoyant particles (\ie{} $d\leq\eta$) follow the flow dynamics and have been used as Lagrangian {\it tracers}~\cite{Mordant:prl2001,Voth:jfm2002,Toschi:ar2009}. But as soon as the particle diameter is of the order of, or greater than the dissipation length scale, the particle dynamics becomes more complicated. Experiments have shown that the dynamics of finite-size particles ($5\leq d/\eta \leq40$) deviates from that of tracers, even when particles are neutrally buoyant. The variance of their acceleration $\overline{a_i^2}$ (\ie{} of the forces acting on them) was indeed found to decrease with size as $d^{-2/3}$~\cite{Voth:jfm2002,Qureshi:prl2007,Brown:prl2009,Volk:jfm2011}.

Very few studies have been carried out for particles with size comparable to the integral length scale~\cite{Zimmermann:prl2011,Klein:mst2013,Cisse:jfm2013}, despite their potential importance for applied situations~\cite{Shew:rsi2007,Gasteuil:prl2007,Zimmermann:ps2013,Zimmermann:njp2013}. It has been shown recently that such particles show a very intermittent dynamics of both translation and rotation. The probability density functions (PDFs) of their linear and angular velocities are almost gaussian, while the PDFs of their acceleration for linear and angular motions reveal wide stretched tails. A coupling between rotation and translation develops, consistent with a Magnus (or lift) force~\cite{Zimmermann:prl2011}.

We built on previous work and address the issue of the motion of large particles, setting aside the particle orientation and focusing only on trajectories. We consider in particular the influence of the large scale (inhomogeneous) flow structure and its impact on the particle's dynamics. It complements recent studies~\cite{Klein:mst2013,Cisse:jfm2013} which have focused on the local flow surrounding the particles, but ignoring the influence of anisotropy and inhomogeneity of the flow at large scales. Our motivation stems from the possibility that particles with sizes close to the integral scale may be sensitive to the kinetic energy injection details and hence the large scale structure of the flow. 

Our experiment is performed in a \karman{} flow (described in section~\ref{sec:setup}), in which large particles are tracked optically. We show that as particle diameters are increased to a fraction of the integral scale, they explore non-uniformly the flow volume: there is preferential sampling of specific flow regions (section~\ref{section:sampling}). We then investigate Eulerian flow maps, as captured by the particles motions (section \ref{sec:results}) and connect with small scale Lagrangian data (section \ref{sec:link}) before discussing in detail features of the preferential sampling effect which is the main finding of this study (section \ref{sec:concl}).

\section{Experimental setup}\label{sec:setup}

\subsection{Turbulence generation}\label{sec:turbulence}
\begin{figure}[t]
   \centering
   \includegraphics[width=\columnwidth]{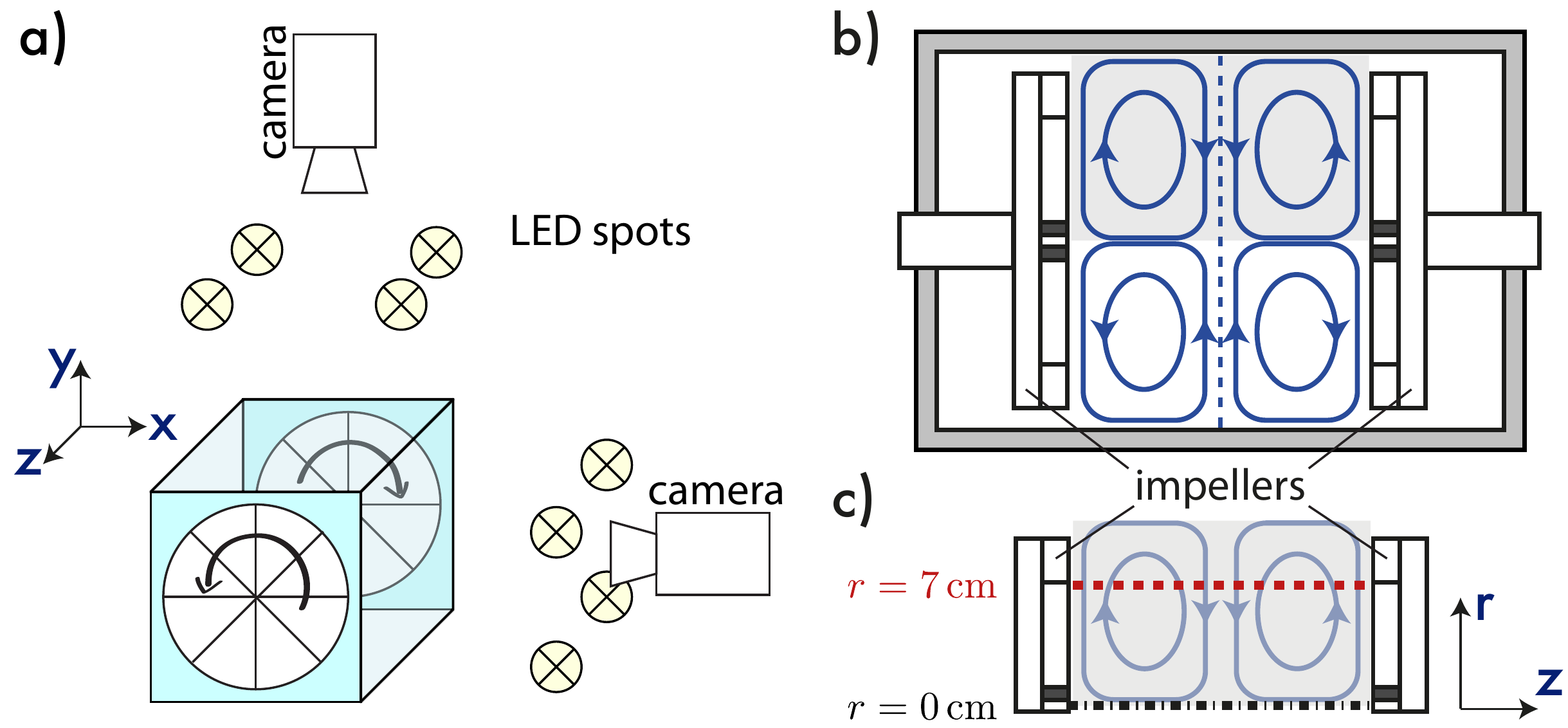} 
   \caption{(a)~Flow vessel and optical setup. (b)~Schematics of the mean flow structure. (c)~Zoom on the half cross section used hereafter to plot maps of data averaged azimuthaly; longitudinal profiles shown later are computed along the red dotted line, at $r=7$~cm.}
   \label{fig:KLAC}
\end{figure}

In this paper, we study the motion of particles in a \karman{} flow, using the same setup as detailed in \cite{Zimmermann:prl2011,Zimmermann:rsi2011}, and summarized in \fig{fig:KLAC}. The turbulent flow is generated in the gap between 2 counter-rotating disks of radius $R=9.5$~cm, fitted with straight blades $1$~cm in height. The flow domain between the disks and walls has a cubic shape with length $H=20\text{ cm}\simeq 2R$ -- the cross section of the vessel is square since flat walls are used to minimize optical distortion. This type of \karman{} swirling flow has been used extensively in the past for the study of {particle dynamics in} fully developed turbulence (see Toschi \textit{et al.} \cite{Toschi:ar2009} and references therein); its local characteristics approximate homogeneous turbulence in its center, although it is anisotropic at large scale~\cite{Ouellette:njp2006, Monchaux:prl2006}. The fluid is set in motion near the two disks in the azimuthal $\theta$-direction, creating two toroidal structures moving in opposite directions, with a strong shear in the center. A poloidal flow is formed as fluid is pumped along the axial $z$-axis towards the center of the disks and an opposite motion occurs along the walls. The combination of the $r$- and $z$-components forms recirculation cells called poloidal structures. It can be noted that the use of a square cross section pins the center shear layer {in the $z=0$ plane (mid plane of the vessel)}  when the two impellers work at the same frequency, contrary to \karman{} flows generated in circular cross section vessels which can exhibit an instability and reversals for the shear layer (\eg{}~\cite{delaTorre:prl2007}).

The working fluid is a water-glycerol mixture, whose density can be finely adjusted to the density of the PolyAmid spheres used here. The resulting mixture has a density of 1.14~g.cm$^{-3}$ and a dynamic viscosity $\mu= 7.5 \cdot 10^{-3}$~Pa.s at working temperature \mbox{$\Theta=20$~\textdegree{}C.} Temperature is controlled by cooling plates at each end of the vessel. The particles are PolyAmid spheres with diameters $d=[6,\,10,\,18,\,24]$~mm (accuracy $0.01$~mm, Marteau \& Lemari\'e, France) and a density $\rho_\text{PA} = 1.14$~g.cm$^{-3}$. The particles are {\it large} in this study, meaning that their diameter is of the order of the integral scale. Table~\ref{table:watergly} summarizes the flow parameters.

\begin{table}[tbh]
\caption{\label{table:watergly}Experimental parameters. $f_\text{imp}$: rotation frequency of the counter rotating disks; $\varepsilon$: energy dissipation estimated from the electrical power consumption of the motors; $Re\equiv (2 \pi R^2f_\text{imp})/\nu$: Reynolds number computed using the disks tip velocity; $R_\lambda=\sqrt{15\cdot 2\pi L_\text{int}^2 f_\text{imp}/\nu}$: Reynolds number based on the Taylor microscale, with $L_\text{int}=3$~cm the estimated integral length scale; $\tau_\eta \equiv (\nu/\varepsilon)^{1/2}$: Kolmogorov time scale; $f_\text{sampling}$: cameras sampling frequency; $\eta \equiv (\nu^3/\varepsilon)^{1/4}$: Kolmogorov length scale;  $d/\eta$: length ratio between the particle diameter and the dissipation scale. The particles diameter ranges between $1/5\, L_\text{int}$ and $4/5\, L_\text{int}$.}
\begin{indented}
\item[]
\begin{tabular}{@{}cccccccc}
\br
$f_\text{imp}$ &  $\varepsilon$ & $Re$ & $R_\lambda$ & $\tau_\eta$ & $f_\text{sampling}$ & $\eta$ & $d/\eta$\\
$\,$[Hz] &  [W.kg$^{-1}$] &  &  & [ms] & [Hz] & [\textmu m]  & \\
\mr
2 &  0.58  & 17\,300 & 161 & 3.4 &    800 & 148   & [40--162]\\
3 &  1.89  & 26\,000 & 197 & 1.9 &  1\,000 & 110   & [54--218]\\
4 &  4.35  & 34\,700 & 228 & 1.2 &  1\,500 &   89   & [67--268]\\
\br
\end{tabular}
\end{indented}
\end{table}

\subsection{Particles tracking}
The motion of the particles is tracked using 2 high-speed video cameras (Phantom V.12, Vision Research, 1Mpix@6kHz) recording synchronously 2 views at approximately 90 degrees. The flow is illuminated by high power LEDs and sequences of 8~bit gray scale images are recorded at a rate $f_\text{sampling}$ ranging from 800 to 3000 frames per second depending on the rotation frequency of the impellers (see Table~\ref{table:watergly}). Both cameras observe the measurement region with a resolution of  $725\times 780$~pixels, covering a volume of $20\times 20\times 15$~cm$^3$ (the shortest distance being between the impellers). Hence,  the particles cover between 20 and 120~pixels depending on their size. The camera are located approximately 2~m away from the apparatus and little variations of the particles size is observed as they move in the flow. Track durations are limited by the cameras on-board memory which stores about $14,000$ frames. Great experimental care is taken for the illumination of the flow so that the particles luminosity is uniform, and shadows and reflections are negligible. 

Particle detection and tracking is carried out using {\sc Matlab}$^\text{\textregistered}$ routines and its image and signal processing toolboxes. For each acquisition, we compute the background as the average of an equally distributed subset of its images. We then subtract the background for each frame and threshold to detect the blobs corresponding to particles.

\begin{figure}[tb] 
   \centering
   \includegraphics[width=\columnwidth]{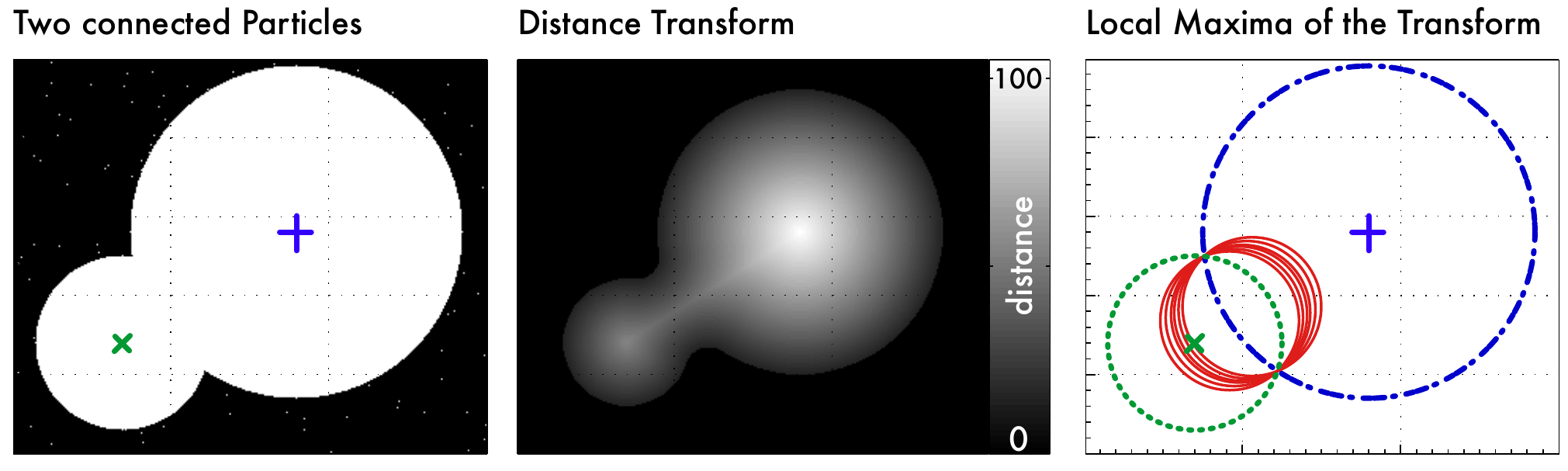} 
   \caption{Sketch of the Blob-splitting technique. Overlapping particles form a blob (left). The distance transform then returns for each white pixel the euclidian distance to the closest black pixel (middle). The local maxima of the distance transform are possible particle positions with their associated radii. Often more maxima than particles  are detected and one has to remove artifacts: starting from the largest radius one  iteratively excludes wrong detections which are within a bigger particle (solid red circles, right). One finally obtains the position $(x,y)$ and radius $r$ of the disentangled particles (dotted green and dashed blue circles).}
   \label{fig:blob}
\end{figure}

For round, unconnected blobs we directly save their location $(x,y)$ on the image together with their diameter (in pixels). 
Connected blobs are split using the maxima of the distance transform~\cite{Soille:book2003} of the blob as sketched in \fig{fig:blob}; the resulting disentangled positions and diameters are then separately stored.  Finally, a standard stereo-matching technique is performed to obtain each particle position in 3D, using Tsai's camera model and calibration technique~\cite{Tsai:ieee1987}. If multiple particles are present in the flow (in order to optimize the experiments and obtain larger statistics), the track assembly is sometimes not straightforward. The main difficulties are removed by the following three steps. 1) Sort the detected particles by diameter. 2) Connect particles of identical sizes using a nearest neighbor algorithm with a short interpolation scheme (namely, gaps of less than 7 times the Kolmogorov time scale are interpolated, corresponding to a number of frames between 15 and 20 depending on  turbulence level and frame rate). Due to their large size particles optically disappear when they enter the projected ``shade'' of another particle. We therefore apply the algorithm suggested in \cite{Xu:mst2008} to reconnect tracks; it enforces continuity for both position and velocity. 3) Identify and eliminate outliers with a least-square spline.

\section{Non uniform sampling} \label{section:sampling}
Neutrally buoyant particles, with inertial range sizes ($D \sim [10-20] \eta$), are known not to form clusters in homogeneous turbulence \cite{Fiabane:pre2012}, leading to a homogenous sampling of the flow. However, it has not yet been investigated if this is still true when particles are freely advected in a flow with a mean structure and non homogeneous properties, especially when particles have diameters of the order of the largest flow eddies. This is the question addressed in this section. 

\begin{figure}[tb]
\centerline{\includegraphics[width=.95\textwidth]{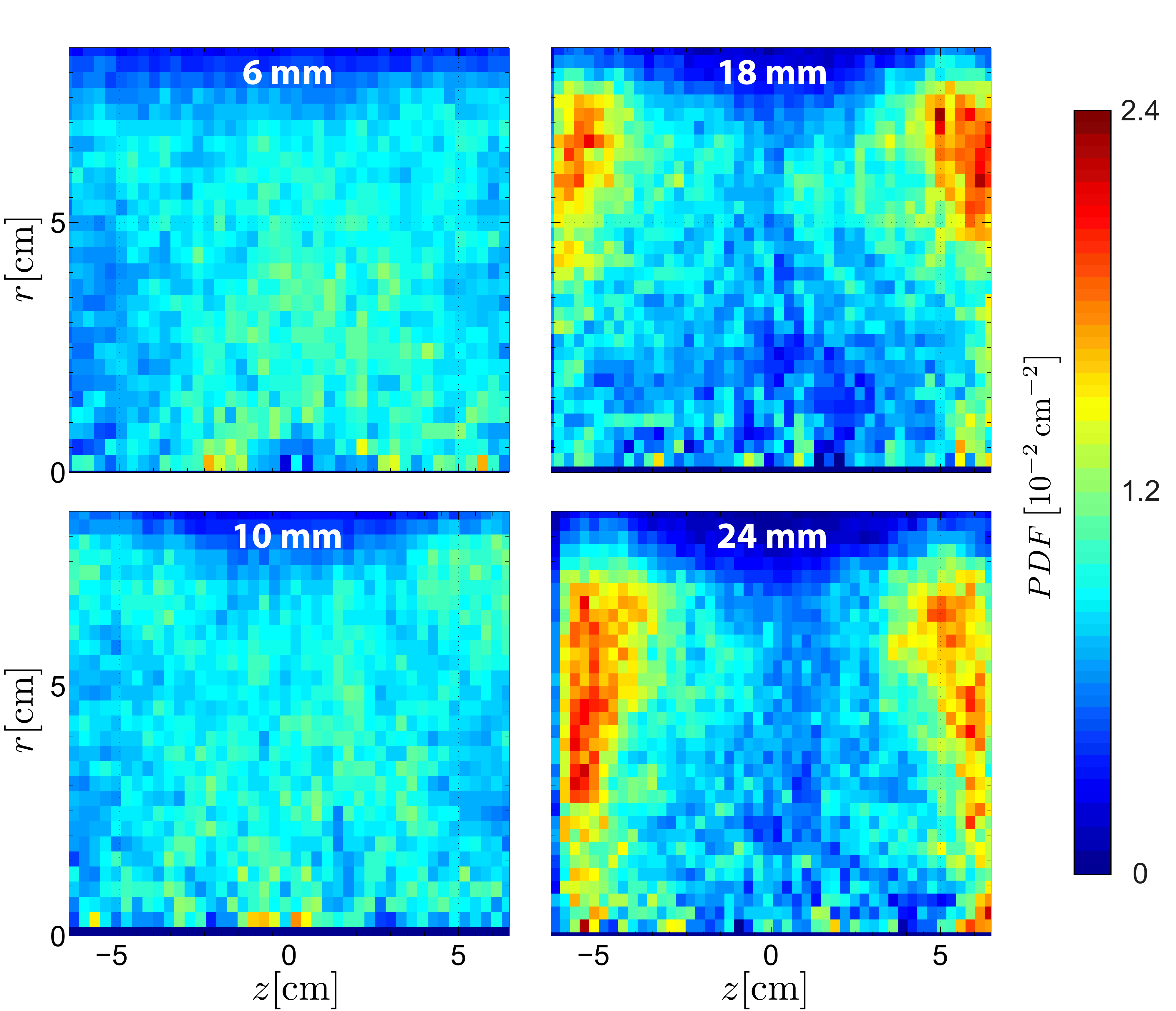}}
\caption{Maps of the probability density functions of position for different particle diameters. On the left-hand side, the flow is homogeneously explored by the smaller particles (6~mm and 10~mm) whereas the larger particles (18~mm and 24~mm) on the right-hand side move preferentially near the impellers. Data from experiments at $f_\text{imp}=4$~Hz.}
\label{fig:pdfpos} 
\end{figure}
 
To answer this question, we first compute the stationary probability density function (PDF) of the particle positions $P(r,\theta,z)$ in cylindrical polar coordinates, and show results for axisymmetric part of the PDFs, noted $\langle P \rangle_{\theta}(r,z)$, in \fig{fig:pdfpos}. Results are displayed  in a half cross section of the vessel, with the impellers on each side (at $z=\pm 10$ cm) and the rotation axis in the center, as sketched in \fig{fig:KLAC}(c).

As clearly shown in \fig{fig:pdfpos}, two very different cases are observed depending on whether the particles are smaller (6~mm and 10~mm, left-hand side in \fig{fig:pdfpos}) or larger (18~mm and 24~mm, right-hand side in \fig{fig:pdfpos}). Particles smaller than $10$ mm explore the flow homogeneously with a probability almost uniform in a meridian plane. Such is not the case for larger particles. Particles with $18$ mm and $24$ mm diameters do not explore the flow uniformly, the probability density of finding the particle with $\|z\|\geq 4$ cm being twice larger than the corresponding value for $\|z\|\leq 4$ cm. On a geometric point of view, analyzing the maps, one finds larger particles avoid the central part of the flow, and move preferentially in two tori with large radius $R_T \simeq 6$ cm, located $5$ cm away from the mid plane, the disks location being at $z=\pm 10$ cm. 

This development of a preferential flow sampling for large particles is a rather unexpected observation. Before further analysis, we need some insight on the properties, Eulerian and Lagrangian, of the carrier flow. In particular, it is of importance to correlate the probability maps with the mean and fluctuating velocity field maps.

\section{Eulerian maps} \label{sec:results}
We discuss here Eulerian flow characteristics, as reconstructed from the Lagrangian motion of the particles, with emphasis on the influence of the particle sizes.

\subsection{Mean velocity field}
The 3D particle tracking yields a set of particle trajectories each containing the temporal evolution of Lagrangian velocity $\mathbf{v}_L$ and acceleration $\mathbf{a}_L$ at the particle position $\mathbf{x}(t)$. Based on this ensemble of trajectories, and for each type of particle, one may define an effective Eulerian flow field $\mathbf{v}_\text{E}(r,\theta,z,t)$ by an Eulerian conditioning of the Lagrangian dataset. Assuming ergodic dynamics, one then obtains a mean velocity field $\overline{\mathbf{v}}_\text{E}\big{(}r,\theta, z\big{)=}\big{(}\overline{v}_r,\overline{v}_\theta,\overline{v}_z\big{)}$ and the rms fluctuations of each velocity component $(v^\text{rms}_r,v^\text{rms}_\theta,v^\text{rms}_z)$.

\begin{figure}[h]
\centerline{\includegraphics[width=\textwidth]{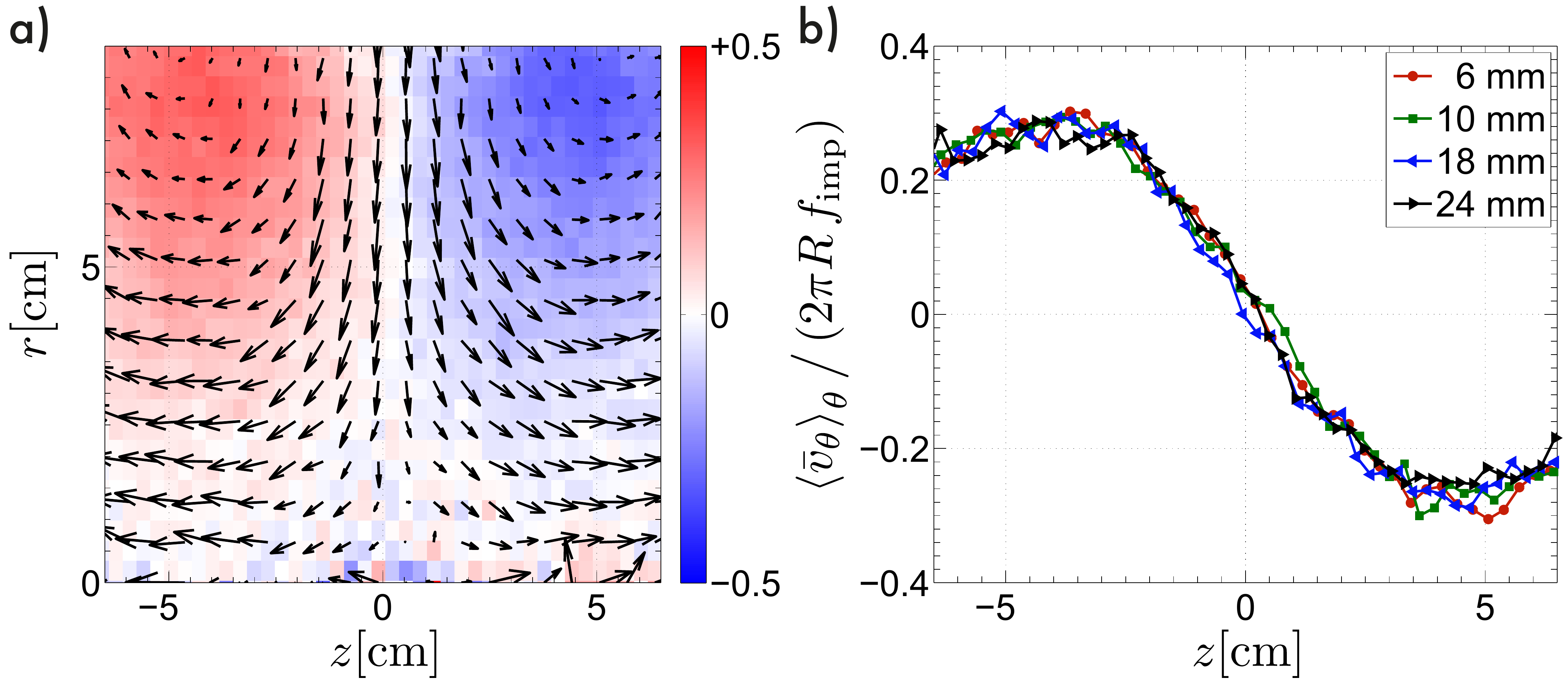}}
\caption{(a) Half cross section of the reconstructed Eulerian mean velocity field normalized by the large scale driving $U=2 \pi Rf_\text{imp}$. Arrows indicate the poloidal $z$- and $r$- components; colors show the azimuthal $\theta$-component. Data obtained with a 6~mm particle at $f_\text{imp}=4$~Hz. (b) Axial profile of the mean azimuthal velocity for particles with diameters between 6 an 24~mm (normalized by the disk tip speed $U=2\pi Rf_\text{imp} \simeq 2.4$ m.s$^{-1}$) at $f_\text{imp}=4$~Hz.}
\label{fig:vel_sections} 
\end{figure}

\fig{fig:vel_sections}(a) shows the map, obtained for the $6$ mm particles, of the axisymmetric part of the reconstructed Eulerian velocity field $\langle \overline{\mathbf{v}}_\text{E} \rangle_\theta (r,z)$. Although $6$ mm particles are not strictly flow tracers, one recognizes in \fig{fig:vel_sections}~(a) the mean structure characteristic of the counter-rotating \karman{} flow: an antisymmetric toroidal component $\langle \overline{\mathbf{v}}_\theta \rangle_\theta$ with respect to the mid plane (colors) and a poloidal flow $(\langle \overline{\mathbf{v}}_r \rangle_\theta, \langle \overline{\mathbf{v}}_z \rangle_\theta)$ showing two recirculation cells (arrows).

In order to study the influence of the particle diameter, we show the evolution of the mean toroidal component $\langle \overline{v}_\theta \rangle_\theta$ as a function of $z$ at a given radius $r_0=7$~cm (along the red dotted line in \fig{fig:KLAC}~©). This location has been chosen such that the mean azimuthal velocity is significantly larger than any possible bias due to measuring incertitudes or lack of statistics. As depicted in \fig{fig:vel_sections}(b), all profiles collapse onto the same curve. This surprising finding is very robust with respect to both velocity components and profile locations, providing the mean velocity is strong enough. It means that neutrally buoyant particles follow the same mean flow independently of their size. Incidentally, it also means that given enough statistics, one may reconstruct the Eulerian mean velocity field from the trajectory of a single neutrally buoyant particle, whatever its diameter!

\subsection{Fluctuating velocity field} \label{sec:discussion_v}
We now consider the velocity fluctuation $v_\text{E}'$ defined as:
\begin{equation}
v_\text{E}'(r,z)=\left( \frac{\langle (v^\text{rms}_r)^2 \rangle_\theta+\langle (v^\text{rms}_\theta)^2 \rangle_\theta+\langle (v^\text{rms}_z)^2 \rangle_\theta}{3}\right)^{1/2}= \left( \di \sum_{i=r,\theta,z} \frac{\langle (v^\text{rms}_i)^2 \rangle_\theta }{3} \right)^{1/2}.
\label{eq:vel_fluct}
\end{equation}

\begin{figure}[h]
\centerline{\includegraphics[width=\textwidth]{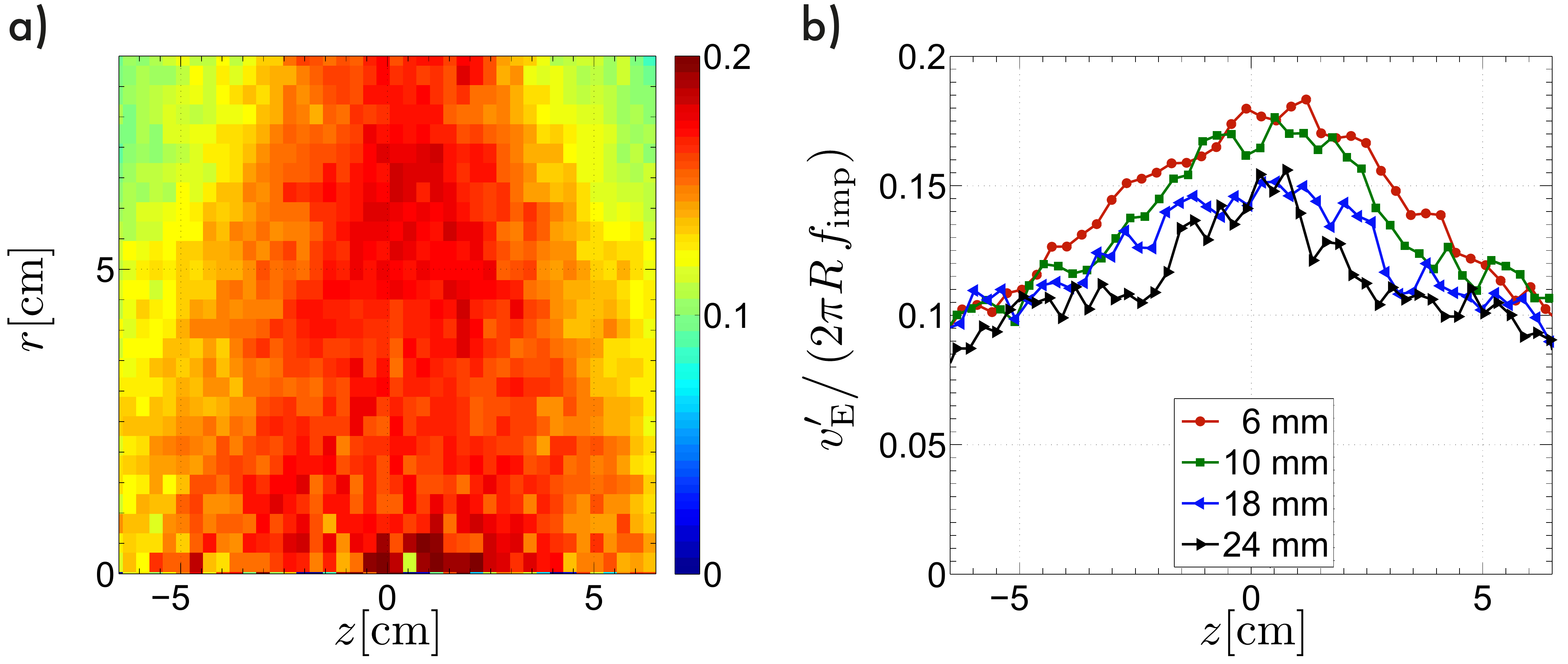}}
\caption{(a)~Half cross section of the normalized fluctuating velocity $v_\text{E}'(r,z) / \left(2 \pi \, R \, f_\text{imp} \right)$. (b) Longitudinal profile at $r=7$~cm for increasing particle diameters. Data from experiments at $f_\text{imp}=4$~Hz.}
\label{fig:profile_vel_fluct} 
\end{figure}

The resulting map and profiles are shown in \fig{fig:profile_vel_fluct}, computed for the smallest particles, with diameter 6~mm. A first observation is that the velocity fluctuations are strongly inhomogeneous. This is consistent with observations made with tracer particles (\eg{} LDV results~\cite{Ravelet:jfm2008}). Fluctuations are larger near the rotation axis, close to the mid plane, and diminish where the toroidal component is the strongest. Again, these results tend to show that $6$ mm particles, while not tracers, have a dynamics that correctly portrays the Eulerian flow.

Studying the longitudinal profiles, one observes in \fig{fig:profile_vel_fluct}~(b) that all have a similar shape, with higher fluctuations in the center (in or near the shear layer) than closer to the impellers. In contrast to the mean velocity, there is a clear sorting of the profiles especially in the central part of the vessel, where the  normalized rms velocity $v'_E/(2 \pi R f_{\text{imp}})$ decreases by over $10\%$ when the particle diameter increases from $6$ mm to $24$ mm. This observation is robust for any location of the profile along the $r$-axis in the volume although it may vary in amplitude across the $(r,z)$ plane. This decrease of the velocity fluctuations with increasing particle diameter differs from what has been reported for particles with diameters close to the dissipation scale~\cite{Qureshi:prl2007,Volk:jfm2011} (for which sizes $d/\eta$ were in the range [1 -- 45] and [12 -- 26] respectively). Although part of those measurements were performed in the central region of the same turbulent flow, no dependence of the rms velocity on the particle diameter was observed. 

As a partial explanation, one may recall that eddies with sizes of the order of the integral scale contain much more kinetic energy than those in the near dissipative range. Thus particles of integral sizes may average fluid fluctuations with a larger impact than is done by particles closer to the dissipation scale.

\subsection{Accelerations} \label{sec:accfuct}
Following the above observations, we now consider the accelerations (at Eulerian fixed points) of the particles. Lagrangian studies have shown that this quantity is quite sensitive to particle sizes \cite{Voth:jfm2002,Qureshi:prl2007,Brown:prl2009,Volk:jfm2011}. We define the acceleration fluctuation:
\begin{equation}
a_\text{E}'(r,z) = \left( \di \sum_{i=r,\theta,z} \frac{\langle (a^\text{rms}_i)^2 \rangle_\theta}{3}\right)^{1/2}.
\label{eq:acc_fluct}
\end{equation}

\begin{figure}[h]
\centerline{\includegraphics[width=\textwidth]{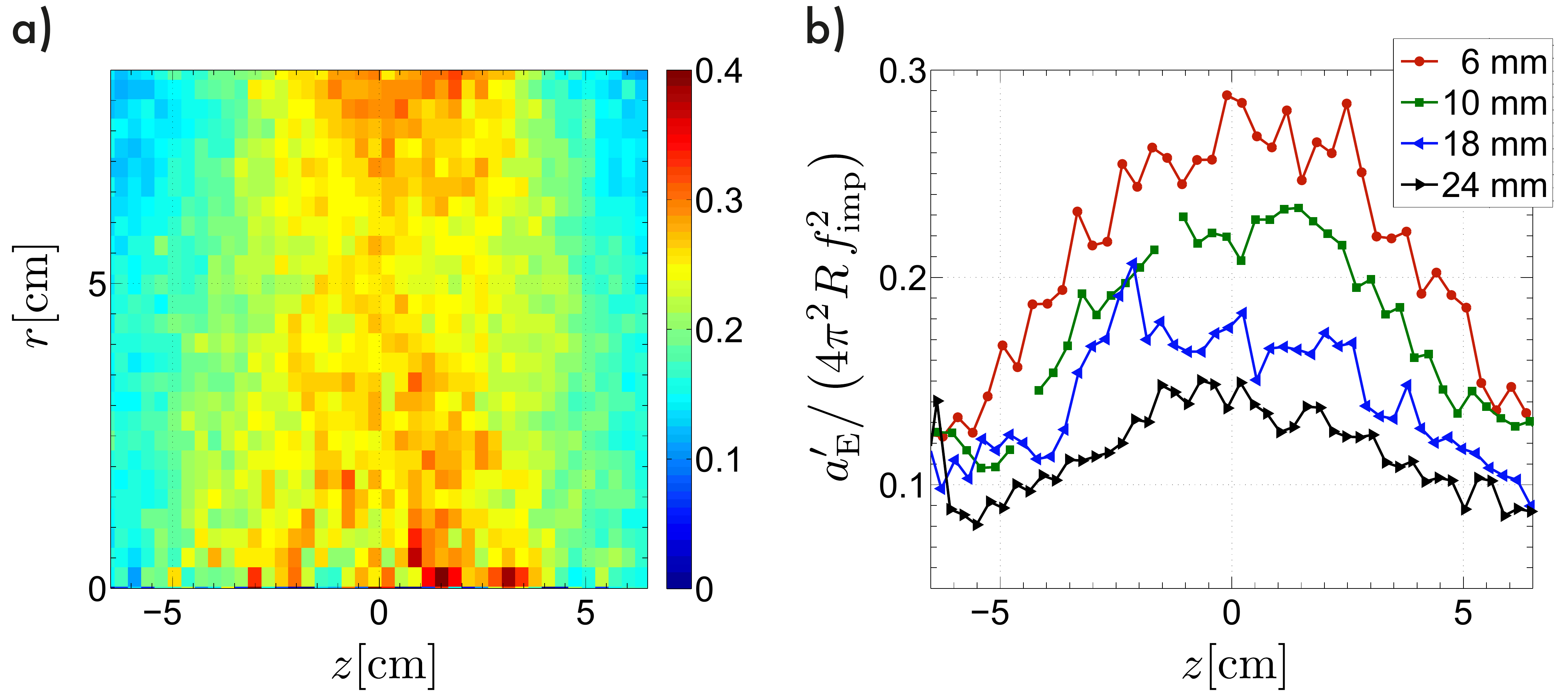}}
\caption{(a)~Half cross section of the normalized fluctuating acceleration $a_\text{E}'(r,z)/\left(4\pi^2 \, R \, f_\text{imp}^2\right)$ for a $6$~mm particle. (b) Longitudinal profile at $r=7$~cm for increasing particle diameters. Data from experiments at $f_\text{imp}=4$~Hz.}
\label{fig:acc_sections} 
\end{figure}

The corresponding maps and profiles are shown in \fig{fig:acc_sections}. One observes that the acceleration fluctuations are much higher near the mid plane of the vessel than to the sides where the mean flow is stronger. This is in agreement with previous observations on velocity, even though high acceleration fluctuations seem to be more localized in the center of the vessel even for small radii.  As already known for smaller particles \cite{Qureshi:prl2007,Brown:prl2009,Volk:jfm2011}, we observe in \fig{fig:acc_sections}~(b) a stronger decrease of acceleration magnitude when increasing the particle diameter than what was observed for velocity fluctuations. Although profiles seem to have a similar shape, $a'_E$ is divided by $2$ when size is increased fourfold. This is a stronger decrease than what was observed for particles of sizes in this inertial range, it will be further investigated together with its Reynolds number dependence in section \ref{sec:link}. We note that the bell shape for the acceleration profile, with a larger value near the mid plane ($z=0$), is similar with local measurements of dissipation $\varepsilon$, as computed from hot wire measurements by Zocchi {\it et al.} in a \karman{} of Helium \cite{Zocchi:pre1994}. It is consistent since Lagrangian accelerations of flow tracers are related to dissipation by $a_{\text{rms}} \sim \varepsilon^{3/4} \nu^{-1/4}$.

\section{Lagrangian quantities} \label{sec:link}
In an inhomogeneous flow, the particles mean and fluctuating quantities in their Lagrangian motion are also of interest, eventually to be connected to the known behavior of Lagrangian {\it tracers}. This may be done in two ways, the first is to compute `standard' Lagrangian estimates using the Lagrangian velocity $\mathbf{v}_\text{L}$ and acceleration $\mathbf{a}_\text{L}$. They can be split into an ensemble average plus a fluctuating quantity ($\mathbf{v'}_\text{L},\mathbf{a'}_\text{L})$ with rms values $(v^{\text{rms}}_{i,\text{L}},a^{\text{rms}}_{i,\text{L}})$,  $i=r,\theta,z$. A second possibility is to use the same Eulerian conditioning as in the previous section, since large particles do not sample the flow uniformly. One then makes use of the position PDFs ($P(r,\theta,z)$) and defines the rms values of velocity (or acceleration) components as :
\begin{equation}
v^{\text{rms}}_{i,\text{LE}}=\left(\iiint \text{d}r\text{d}\theta \text{d}z r P(r,\theta,z) \, (\overline{v}_i(r,\theta,z))^2\right)^{1/2},  \; \; i=r,\theta,z.
\label{eq:unbiased}
\end{equation}
The subscript `LE' emphasizes here that one wishes to estimate the Lagrangian fluctuations, taking into account the non uniformity of the sampling in the Eulerian space \footnote{Technically, rms values are computed by removing the local axisymmetric part of the mean velocity (or acceleration) at the particle position before computing $v^{\text{rms}}_{\text{LE}}$ as a standard deviation of the dataset. It ensures all grid points are weighted by the time particles spent at specific locations in the flow.}. 

\begin{figure}[tb]
\centerline{\includegraphics[width=\textwidth]{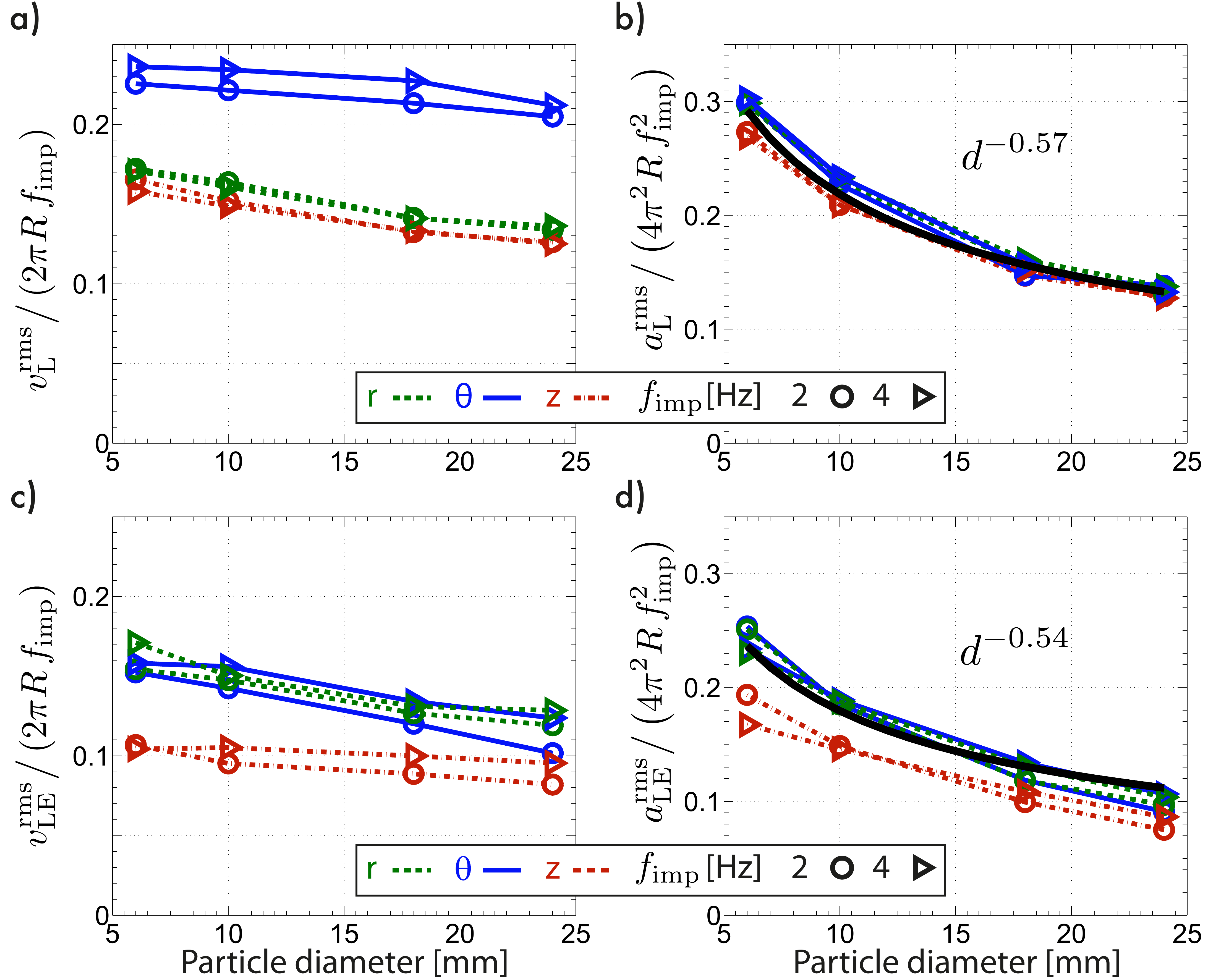}}
\caption{Standard deviation (root mean square values) of normalized quantities as a function of the particle diameter $d$ for two impeller frequencies $f_\text{imp}=2\text{ Hz}$ and $f_\text{imp}=4\text{ Hz}$. (a)~Biased Lagrangian velocity components $v^\text{rms}_i$. (b)~Biased Lagrangian acceleration components $a^\text{rms}_i$ -- plain black line is a power-law fit in $d^{-0.56}$. (c)~Unbiased Lagrangian velocity components $v^\text{rms}_i$ using Eulerian conditioning and position PDFs weighting. (d)~Unbiased Lagrangian acceleration components $a^\text{rms}_i$ using Eulerian conditioning and position PDFs weighting -- plain black line is a power-law fit in $d^{-0.54}$.}
\label{fig:varms} 
\end{figure}

Results for the normalized Lagrangian rms velocity $v^\text{rms}_\text{L}/(2 \pi R f_{\text{imp}})$ and acceleration  $a^\text{rms}_\text{L}/(4 \pi^2 R f_{\text{imp}}^2)$ are displayed in \fig{fig:varms}(a,b), together with the corresponding normalized Eulerian velocity $v^\text{rms}_\text{LE}/(2 \pi R f_{\text{imp}})$ and acceleration  $a^\text{rms}_\text{LE}/(4 \pi^2 R f_{\text{imp}}^2)$ in \fig{fig:varms}(c,d). This normalization collapses the velocity measurements made at different rotation rates of the driving disks. It is characteristic of a fully turbulent regime for which velocity mean and fluctuating quantities are proportional to the large scale forcing~\cite{Ravelet:jfm2008}. 

However, anisotropy and characteristic values differ depending on the way estimates are computed. Indeed, the direct Lagrangian estimate $v^{\text{rms}}_\text{L}$, for which ensemble averaged velocity reduces to zero in a bounded flow, incorporates mean flow contribution in the fluctuations. One then finds $v^{\text{rms}}_{i,\text{L}}$ is always larger than its corresponding component $v^{\text{rms}}_{i,\text{LE}}$. This is particularly clear for the toroidal component $v^{\text{rms}}_{\theta,\text{L}}$, found to be much larger than any other velocity component because of the added contribution from the mean azimuthal velocity, an antisymmetric quantity with respect to $z$ coordinate. When the mean flow contribution is removed in \fig{fig:varms}~(c), the measured anisotropy is in better agreement with previous observations in \karman{} flows~\cite{Voth:jfm2002,Ravelet:jfm2008}. We also note that the anisotropy seems to decrease with increasing particle size.

Contrary to velocity fluctuations, acceleration fluctuations -- \fig{fig:varms}(b,d) -- are very similar for all components and there is less influence from the methods used for computing the estimated rms (L or LE). This reduction of anisotropy for small scale quantities such as the acceleration may be expected, and is in agreement with measurements reported in the literature (\eg~\cite{Voth:jfm2002,Ouellette:njp2006}) for Lagrangian {\it tracers}.  Concerning the evolution with Reynolds number, one finds particle acceleration is correctly normalized by the large scale acceleration $4 \pi^2 R f_{\text{imp}}^2$. This differs from the scaling $a^{\text{rms}} \sim \varepsilon^{3/4} \nu^{-1/4}$ observed for Lagrangian tracer particles in the same flow \cite{Brown:prl2009}. It indicates that large particles no longer feel the dissipative scale $\eta$, and have an acceleration almost independent of the viscosity. As already observed in section \ref{sec:accfuct}, one can see in \fig{fig:varms}(b,d) that rms acceleration decreases more rapidly with increasing particle size than the rms velocity. When power laws are tested, one finds a behavior compatible with  ${a^\text{rms}(d)}/{f_\text{imp}^2} \propto d^\alpha$ and a scaling exponent close to $\alpha \simeq -0.5$.  This is a larger variation than $\alpha \simeq -1/3$ observed for smaller particles~\cite{Qureshi:prl2007,Volk:jfm2011}. Note that as the dissipation is non homogeneous in space, one may expect the exponent to depend on space so that the value $\alpha \simeq -0.5$ should be considered as a global estimate of the acceleration magnitude decrease. We also note that the measured exponent found in \cite{Volk:jfm2011} was already larger than the one found in wind tunnel experiment \cite{Qureshi:prl2007}, with a difference that can originate from the mean structure of the \karman{} flow.

\section{Discussion and conclusion remarks}\label{sec:concl}
The different maps, of position PDF, or those of velocity and acceleration fluctuations show the strong impact of the particle size on its dynamics. One can note the interesting correlation between position PDFs displayed in \fig{fig:pdfpos} and the maps of velocity fluctuations of \fig{fig:profile_vel_fluct}(a). Particles with diameters 18~mm and 24~mm, have position PDF maps qualitatively very similar to the color-inverted map of velocity fluctuations. This indicates that larger particles go preferentially in less active regions of the flow with smaller velocity and acceleration fluctuations. These results are in good agreement with turbophoresis, that is the tendency for particles to migrate in the direction of decreasing turbulence level. However, particles are not only affected by velocity fluctuations, but also by the mean flow itself from which a trapping due to the mean pressure gradient may originate. An explanation for the preferential sampling of large particles could be that the decrease of velocity fluctuations for large particles is not sufficient to overcome the trapping from the mean flow while smaller particles are less sensitive to the trapping as compared to velocity fluctuations.

In order to investigate further the origin of this preferential sampling, we have varied many parameters concerning the flow (viscosity, amplitude of large scale forcing) as well as the particles characteristics (size, density, and surface roughness). All these studies have confirmed the robustness of the preferential sampling effect, with respect to parameter variations. No variation of the position PDFs has been observed, for a given size, when Reynolds number $Re=2 \pi R^2 f_{\text{imp}} /\nu$ has been varied by changing the disks velocity. This observation is consistent with the fact that, as reported in section \ref{sec:link}, velocities and accelerations scale with the driving of the flow. Indeed, the relative intensity of trapping as compared to fluctuations is not modified in such a regime. This has been confirmed by recent experiments with increased viscosity $\mu=135$ $\mu_{water}$ for which velocity fluctuations are decreased as compared to the mean flow \cite{Ravelet:jfm2008}. In such a regime, trapping was found to increase on the sides of the measurement volume, near the flow poloidal neutral lines.

In order to investigate and possibly modify the rotation-translation coupling of the particle found in \cite{Zimmermann:prl2011}, the particle surface roughness was also varied and was shown not to modify the position PDFs.

Finally, several other experiments were made with similar particles (spheres of diameters \mbox{$d=[6,\,10,\,18,\,24]$~mm)} while changing the density with respect to the fluid. Using water (unit density), experiments where performed using the same PolyAmid particles with density $\rho_\text{PA}=1.14$~g.cm$^{-3}$, and lighter PolyPropylene particles with a density $\rho_\text{PP}=0.9$~g.cm$^{-3}$. These experiments also allowed to investigate the influence of the viscosity of the fluid ($\mu= 7.5 \cdot 10^{-3}$~Pa.s for the water-glycerol mixture versus $\mu= 1.0 \cdot 10^{-3}$~Pa.s for the distilled water). It was found that the relative density of the particle modified position PDF maps, lighter particles being more efficiently trapped, and heavier particles being less trapped than neutrally buoyant ones of the same diameter. This can be understood in the framework of beta-Stokes model of inertial particles dynamics \cite{calzavarini:2008a}, for which the influence of mean pressure gradient is given by the reduced density ratio $\beta=3 \rho_f/(2 \rho_p +\rho_f)$ through an added mass force. This means that the average pressure gradient exerted on the solid particles, together with velocity fluctuations felt by the particles are the key parameters to understand this preferential sampling effect. In the case of the turbulent \karman{} flow considered here, the preferential sampling occurs around the same axial positions as the flow poloidal neutral flow lines. These structures are stable attractors of the laminar  \karman{} flow. It suggests that freely advected particles, when large enough compared to the dissipation scale, perform some average of the flow field and reveal its low dimensional dynamics. \\

\paragraph {\bf Acknowledgments} The authors want to thank Haitao Xu, Alain Pumir and Javier Burguete for many fruitful discussions, and the machine shop at ENS de Lyon for manufacturing the vessel. This work is part of the International Collaboration for Turbulence Research.
It is supported by French Research Program ANR-12-BS09-0011 ``TEC2'' and European MPNS COST Action MP0806 ``Particles in Turbulence''.

\section*{References}
\bibliography{biblio_Lsphere}

\end{document}